%% file: 4NCOSMO.tex
\documentclass[11pt]{article}

\usepackage{a4,amsmath,amssymb,amscd,graphicx,epsfig,xypic}
\usepackage{vmargin}

\everymath{\displaystyle}

\newtheorem{teo}{Theorem}[section]

\newtheorem{cor}[teo]{Corollary}

\newtheorem{prop}[teo]{Proposition}
\newtheorem{lem-defi}[teo]{Lemma-Definition}
 \newtheorem{defi}[teo]{Definition}

\newtheorem{conge}[teo]{Conjecture}
\newtheorem{remark}[teo]{Remark}

\newcommand{\mr}{\mathbb{R}}

\newcommand{\mz}{\mathbb{Z}}
\newcommand{\mh}{\mathbb{H}}

\newcommand{\mm}{\mathbb{M}}
\newcommand{\mi}{\mathbb{I}}
\newcommand{\mx}{\mathbb{X}}

\title{Cosmological Time in (2+1)-Gravity}

\author{Riccardo Benedetti and Enore Guadagnini}

\date{}
\begin{document}

\maketitle
\vspace{0.3cm}
\noindent Dipartimento di Matematica, Universit\`a di Pisa,
Via F. Buonarroti, 2, I-56127 PISA

\noindent Email: benedett@dm.unipi.it

\vspace{0.3cm}
\noindent Dipartimento di Fisica, Universit\`a di Pisa,
Via F. Buonarroti, 2, I-56127 PISA

\noindent Email:  guada@df.unipi.it
\vspace{1cm}

\input 4NCOSMOINT.tex

\input 4NCOSMOCT.tex

\input 4NCOSMOGST.tex

\input 4NCOSMO3CT.tex

\input 4NCOSMOCOM.tex

\input 4NCOSMOBIB.tex
\end{document}

%% file: 4NCOSMOINT.tex
 \begin{abstract}
\noindent We consider maximal globally hyperbolic flat (2+1)
spacetimes with compact space $S$ of genus $g>1$. For any spacetime
$M$ of this type, the length of time that the events have been in
existence is $M$ defines a global time, called the cosmological time
CT of $M$, which reveals deep intrinsic properties of spacetime.  In
particular, the past/future asymptotic states of the cosmological
time recover and decouple the linear and the translational parts of
the $ISO(2,1)$-valued holonomy of the flat spacetime. The initial
singularity can be interpreted as an isometric action of the
fundamental group of $S$ on a suitable real tree.  The initial
singularity faithfully manifests itself as a lack of smoothness of the
embedding of the CT level surfaces into the spacetime $M$. The
cosmological time determines a real analytic curve in the
Teichm\"uller space of Riemann surfaces of genus $g$, which connects
an interior point (associated to the linear part of the holonomy) with
a point on Thurston's natural boundary (associated to the initial
singularity).
\end{abstract}
\bigskip

\smallskip

\section{Introduction.}

\noindent We shall be mainly concerned with maximal globally
hyperbolic, matter-free spacetimes $M$ of topological type $S \times
\mr$, where $S$ is a compact closed oriented surface of genus
$g>1$. The (2+1)-dimensional Einstein equation with vanishing
cosmological constant actually implies that $M$ is (Riemann) flat.

\noindent After \cite{D-J-'tH} and \cite{W}, a large amount
of literature has grown up about this $(2+1)$-gravity topic,
regarded as a useful toy-model for the  higher
dimensional case. Two main kinds of description have been
experimented. A ``cosmological'' approach  points to
characterize the spacetimes in terms of some distinguished global
time; for instance the {\it constant mean curvature} CMC time
has been widely studied \cite{A-M-T}, \cite{Mo}. A
``geometric'' time-free approach eventually identifies each
flat spacetime by means of its $ISO(2,1)$-valued holonomy
\cite{W}, \cite{Me}.
With the exception of the case with toric space ($g=1$),
there is not  a clear correspondence
between the results obtained in these two approaches.

\noindent The aim of this paper is to show that this gap can be filled
by using the canonical Cosmological Time CT, that is ``the length of
time that the events of $M$ have been in existence'' (see
\cite{A-G-H}). It turns out that this is a global time which reveals
the fundamental properties of spacetime.  It is canonically defined by
means of the very basic spacetime's structures: its casual structure
and the Lorentz distance. The cosmological time $\tau$ is invariant
under diffeomorphisms, therefore the $\tau =a$ level surfaces $S_a$
provide a gauge-invariant description of space evolution in $M$.  Both
the intrinsic and extrinsic geometry of the surfaces $S_a$, as well as
their past/future asymptotic states, are intrinsic features of
spacetime. The asymptotic states are defined by the evolution of the
observables associated to the length of closed geodesic curves on the
surfaces $S_a$. Remarkably, they recover and decouple the linear and
the translational parts of the holonomy.  The study of the asymptotic
states also leads to understand the initial singularity (we will
always assume that the space is future expanding) and the way how the
classical geometry degenerates, but does not completely disappear.
The initial singularity can be interpreted as the isometric action of
the fundamental group of $S$ on a suitable ``real tree''. Differently
to the case of the CMC time (for instance), the level surfaces $S_a$
of the CT are in general only $C^1$-embedded into the spacetime
$M$. This lack of smoothness takes place on a ``geodesic lamination''
on $S_a$ and is a observable large scale manifestation of the
intrinsic geometry of the initial singularity.  Thus the initial
singularity admits two complementary descriptions: one, in terms of
real trees and, the second, in terms of geodesic laminations.  The
existence of a duality relation between real trees and laminations was
already known in the context of Thurston theory of the boundary of the
Teichm\"uller space. It is remarkable that Einstein theory of
(2+1)-gravity sheds new light on this subject and puts duality in
concrete form.

\noindent In \cite{BG2} we have also used the cosmological time in order to
study certain interesting families of $(2+1)$-spacetimes coupled to
particles.
\medskip

\noindent Our main purpose consists of elucidating the central role of the
cosmological time and its asymptotic states in the description of
spacetimes. The cosmological time
perspective provides a new interpretation of several facts spread in
the literature which are related to Thurston work. More precisely, the present
article is based on, and could be considered a complement of, Mess's
fundamental paper \cite{Me}.

%% file: 4NCOSMOCT.tex
\section{The Cosmological Time Function.}\label{CTF}

\noindent For the basic notions of Lorentzian geometry and
causality we refer for instance to \cite{B-E}, \cite{H-E}.
Let $N$ be any time oriented Lorentzian manifold of
dimension $n+1$. The {\it cosmological time function},
$\tau: N \to (0,\infty]$, is defined as follows. Let $C^-(q)$ be the
set of past-directed causal curves in $N$ that start at $q\in N$, then

$$ \tau(q) = {\rm sup}\, \{L(c):\ c\in C^-(q)\} $$

\noindent where $L(c)$ denotes  the Lorentzian length of the curve $c\, $: 
$$ L(c)= \int_c\ (\, {\rm proper-time}\, )\ \ .$$

\noindent 
$\tau(q)$ can be interpreted as  the length of time the event $q$ has been in
existence in $N$. For
example, if $N$ is the standard flat Minkowski space $\mm^{n+1}$,
$\tau$ is the constant $\infty$-valued function, so in this case it is not very
interesting.  In \cite{A-G-H} (see also \cite{W-Y}) one studies the
properties of a manifold $N$ with {\it regular} cosmological time
function. Recall that $\tau$ is regular if:

1) $\tau(q)$ is finite valued for every $q\in N$;

2) $\tau \to 0$ along every past directed inextensible causal
curve.
\smallskip

\noindent The existence of a regular cosmological time function has strong
consequences on the structure of $N$ and of the constant-$\tau$ surfaces
\cite{A-G-H}. In particular when  $\tau$ is regular, 
$\tau  : N\to (0,\infty[$ is
a continuous function, which is twice differentiable almost
everywhere, giving a global time on $N$ denoted by CT. Each $\tau$
level surface is a future Cauchy surface, so that $N$ is globally
hyperbolic. For each $q \in N$ there exists a future-directed
time-like unit speed geodesic ray $\gamma_q: (0,\tau(q)]\to N$ such
that:

$$ \gamma_q(\tau(q)) = q \qquad ,  \qquad \tau(\gamma_q(t)) = t \quad .$$

\noindent The union of the past asymptotic end-points of these
rays can be regarded as the initial singularity of $N$.
\smallskip

\noindent The cosmological time function is not related to any specific
choice of coordinates in $N$; it is ``gauge-invariant'' and so it represents
an intrinsic feature of spacetime. Thus, when the cosmological time is
regular, the $\tau$-constant level surfaces and their properties have
a direct physical meaning as they are observables.

\noindent We present now two basic examples of spacetime with
regular cosmological time, which shall be important throughout all the paper.
To fix the notations, the standard Minkowski space $\mm^{2+1}$ is 
endowed with coordinates $\, x = (x^1,x^2,x^3)$, so that the metric is
given by $ds^2=(dx^1)^2 + (dx^2)^2 -(dx^3)^2$.  $\mm^{2+1}$  is oriented and
time-oriented in the usual way.
\smallskip

\noindent {\bf Example 1.} Consider the chronological future of the origin 
$0\in \mm^{2+1}$

$$I^+(0) = \{x \in \mm^{2+1}: (x^1)^2 +(x^2)^2 - (x^3)^2 < 0,\
x^3>0\}. $$
\noindent Its cosmological time, $\tau: I^+(0)\to
(0,\infty)$, is a smooth submersion; the constant-time  $\{\tau = a\}$ 
surfaces are the (upper) hyperboloids

$$ \mi (a) = \{x \in \mm^{2+1}: (x^1)^2
+ (x^2)^2 - (x^3)^2 = -a^2,\ x^3>0\}. $$

\noindent Hence $\mi (a)$ is a complete space of constant Gaussian
curvature equal to $-1/a^2$, and of constant extrinsic mean curvature
$1/a$.  The Lorentzian length of the time-like geodesic arc connecting
any $p\in I^+(0)$ with $0$ equals $\tau(p)$; $0$ is the initial
singularity. Note that $\mi(a)$ can be obtained from $\mi (1)$ by
means of a dilatation in $\mm^{2+1}$ with constant factor $a$; 
shortly we write $\mi(a)=a\mi (1)$.  We shall denote by $SO(2,1) $
the group of oriented Lorentz transformations acting on
$\mm^{2+1}$ and by $ISO(2,1) $ the Poincar\'e group.  $SO^+(2,1)$
denotes the subgroup of $SO(2,1)$ transformations which keep
$I^+(0)$ and each $\mi(a)$ invariant.  $ISO^+(2,1) $ is the corresponding
subgroup of $ISO(2,1)$.
   
\smallskip

\noindent {\bf Example 2.} Let us denote by $I^+(1,3)$
the chronological future in $\mm^{2+1}$ of the line 
$\{\ x^1 = x^3 =0\ \}$
$$I^+(1,3) =  \{x \in \mm^{2+1}: (x^1)^2 - (x^3)^2 <
0,\ x^3>0\} \ . $$

\noindent The Lorentzian length of the
time-like geodesic arc connecting any $p=(x^1,x^2,x^3)\in I^+(1,3)
$ with $q=(0,x^2,0)$ equals the cosmological time $\tau(p)$. 
The level surfaces are

$$\mi(1,3,a)= \{x \in \mm^{2+1}: (x^1)^2 - (x^3)^2 = -a^2,\
x^3>0\}  $$

\noindent and have constant extrinsic mean curvature equal to
$(1/2a)$.  Each surface  $\mi(1,3,a)$ is isometric to
the flat plane $\mr^2$. To make this manifest, it is useful to
consider the following change of coordinates.  Let $\Pi^{2+1} = \{
(u,y,\tau)\in \mr^{2+1}: \tau >0\}$ be  endowed with the metric       
  $ds^2= \tau^2du^2 + dy^2  - d\tau^2$.
Then, $ x^1 = \tau sh(u),\  x^2 = y,\ x^3 = \tau ch(u) $, is an
isometry between $\Pi^{2+1}$ and  $I^+(1,3)$. The level set
$\{\tau = a\}$ of  $\Pi^{2+1} $ goes isometrically  onto  $\mi(1,3,a)$, so this
is intrinsically flat.  Note that the group of oriented isometries
of $\Pi^{2+1}$ is generated by the translations parallel to the
planes $\{ \tau = a\}$, and the rotation of angle $\pi$ of the $(u,y)$ 
coordinates.
\medskip

\noindent We are going to show that any maximal globally
hyperbolic, matter-free  $(2+1)$-spacetime $M$, with compact
space $S$, actually has regular cosmological time, and its initial
singularity can be accurately described.

%% file: 4NCOSMOGST.tex
\section{Flat (2+1)-spacetimes}\label{GST}

A flat spacetime is, by definition, locally isometric to the
Minkowski space $\mm^{2+1}$. We assume that our maximal hyperbolic
flat spacetimes are time-oriented and future expanding, and that these
orientations locally agree with the usual ones on $\mm^{2+1}$. The
spacetime structures on $S\times \mr$ are regarded up to oriented
isometry homotopic to the identity.
\smallskip

\subsection{Minkowskian suspensions}\label{suspension}
\noindent We introduce here the simplest $(2+1)$ spacetimes with
compact space $S$ of genus $g>1$.

\begin{figure}
\begin{center}
\includegraphics[width=7cm]{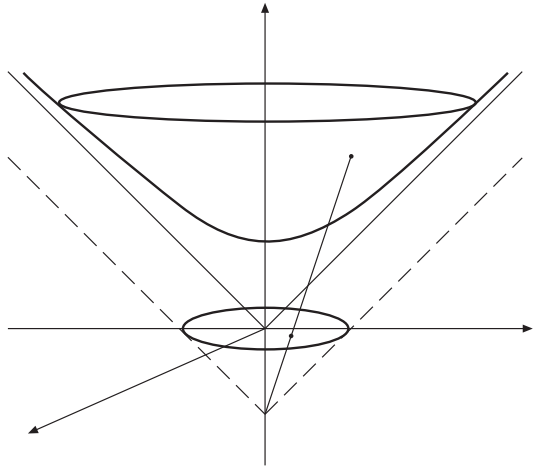}
\caption{The hyperbolic plane in the hyperboloid and disk models.}
\label{fig:H2}
\end{center}
\end{figure}

\noindent Recall that the upper hyperboloid $\mi (1)\subset
\mm^{2+1}$, mentioned in the previous section, is a classical
model for the hyperbolic plane $\mh^2$ (see \cite{B-P} for this and
other models); the Poincar\'e disk is another model which can be
obtained from $\mi (1)$ by means of the stereographic projection shown in
figure \ref{fig:H2}. We shall use the Poincar\'e model in section 4.

\noindent Take any hyperbolic surface $F=\mh^2/\Gamma$
homeomorphic to $S$. $\Gamma$ is a subgroup of $SO^+(2,1)$ which acts
freely and properly discontinuously on $\mh^2 \cong \mi (1)$. $\mi
(1)$ can be identified with the universal covering of $S$ and $\Gamma$
with the fundamental group $\pi_1(S)$. $\Gamma$ can be thought also as
a group of isometries of the spacetime $I^+(0)$ and $M(F) =
I^+(0)/\Gamma$ is the required spacetime with compact space
homeomorphic to $S$. We call it the {\it Minkowskian suspension} of
$F$. This construction is well-known;  sometimes
$M(F)$ is also called the Lorentzian cone over $F$ or the L\"obell
spacetime based on $F$. $I^+(0)$ can be regarded as the universal
covering of $M(F)$.  Let us now consider the cosmological time of $M(F)$. 
The CT of $I^+(0)$ naturally induces the CT of $M(F)$. Indeed, each level
surface $S_a$ of $M(F)$ has $\mi (a)$ as universal
covering; moreover, $S_1 = F$ and  $S_a = aF$. In this case, the CT coincides 
with the CMC time and each level surface  $S_a$ smoothly embeds into
$M(F)$. The initial singularity ``trivially'' consists of one point.
\smallskip

\noindent {\bf Notation.} Let $Y$ be any subset of $\mi (1)$, we shall denote
by $\widehat{Y}$ its
``suspension'' in $I^+(0)$ which is defined by
$\widehat{Y} = \cup_{a\in (0,\infty[}\ aY$.

\subsection { (2+1) spacetimes as deformed Minkowskian suspensions}
\noindent It has been shown in \cite{Me} that  any maximal globally 
hyperbolic, future expanding flat spacetime $M$ with compact space
homeomorphic to $S$, as above, can be regarded as a ``deformation''
of some Minkowskian suspension (see also \cite{W}). In fact $M$ is of the
form $M=U(M)/\Gamma'$, where:

1) The domain $U(M)$ of $\mm^{2+1}$ is a convex set 

$$U(M)= \{x\in \mm^{2+1}; x^3> f(x)\}$$

\noindent where $f: \{x^3=0\}\to [0,\infty[$ is a convex function.

2) $\Gamma'$ is a subgroup of $ISO^+(2,1)$ (also called the {\it holonomy}
group of $M$) acting freely and
properly discontinuously on $U(M)$. Hence $U(M)$ is
the universal covering of $M$ and $\Gamma'$ is isomorphic to
$\pi_1(M)\cong \pi_1(S)$.

3) The ``linear part'' $\Gamma$ of $\Gamma'$ is a subgroup of
$SO^+(2,1)$ which is isomorphic to $\pi_1(S)$ and acts freely and
properly discontinuously on $\mi (1)\cong \mh^2$. This is a non
trivial fact which follows from a result of Goldman \cite{Go}.
Each element $\gamma' \in \Gamma'$ is of the form $\gamma' =
\gamma + t(\gamma)$, where $\gamma \in \Gamma$ and $t(\gamma)\in
\mr^3$ is a translation. $t: \Gamma \to \mr^3$ is a {\it cocycle}
representing an element of $H^1(\Gamma,\mr^3)$. If $t'= \lambda
t$, $\lambda \in \mr^*$, then $U(M')$ differs from  $U(M)$
by : $U(M)=\lambda^{-1} U(M')$. When $\lambda$ is ``small'',
$U(M')$ is ``close'' to $I^+(0)$ ($M'$ is ``close'' to $M(F)$,
$F=\mh^2/\Gamma$).

\noindent Note that $\Gamma'$, whence $U(M)$ and $t$, are defined up
to inner automorphism of $ISO^+(2,1)$.

\subsection {Spacetimes of simplicial type}

\noindent In this section, we shall consider the flat spacetimes that can be 
obtained from Minkowskian suspensions by means of particular
deformations.  These spacetimes will be called of {\it simplicial
type}, the origin of this name is related to the material presented in
section 4. Spacetimes of simplicial type are important because
they are ``dense'' in the set of all spacetimes; the
shape of any spacetime and of its CT can be arbitrarily well
approximated by some spacetime of simplicial type (see
proposition \ref{dense}). So, it is enough to understand these
examples in order to have a rather complete qualitative picture of our
general presentation. Moreover, all the statements of this paper can
easily be checked in a spacetime of simplicial type.
\smallskip

\noindent Start with a Minkowskian
suspension $M(F)$. Assume that a {\it weighted
multi-curve} $\mathcal{L}$ on $F$ is given. $\mathcal{L}$ is the union of a
finite number of disjoint simple closed geodesics on $F$, each one
endowed with a strictly positive real weight. $\mathcal{L}$ governs a
specific deformation of $M(F)$ producing a required flat spacetime denoted by
$M(F,\mathcal{L})$.  A particular spacetime deformation is associated to each
 component of $\mathcal{L}$ and can be obtained by means of an appropriate 
surgery operation in Minkowski space. As the deformations associated to the 
components of $\mathcal{L}$ act locally and independently from each other, 
we may assume  for simplicity  that $\mathcal{L}$ just consists of one 
component $c$, with weight $r$ and length $s$.
\smallskip

\begin{figure}
\begin{center}
\includegraphics[width=4cm]{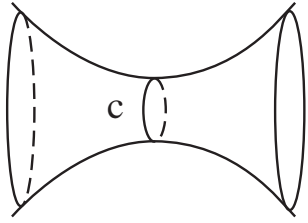}
\caption{The surface $F_0$ with the closed geodesic $c$.}
\label{fig:tromba}
\end{center}
\end{figure}

\noindent {\bf Elementary deformation.} In order to illustrate the 
deformation associated with one geodesic $ c $ with weight $ r $, we shall 
now introduce a simple hyperbolic surface  $F_0$ which can be understood as a 
local model for the general  surface $ S$.  Let
$\Gamma_0$ be an infinite cyclic subgroup of $SO^+(2,1)$ generated
by an element $g_0$ acting on $\mi (1)$ as an isometry of {\it
hyperbolic type} (see for instance \cite{B-P} for the
classification of the isometries of $\mh^2$). We can assume that 
$g_0$ is a Lorentz transformation corresponding to a boost along the 
$x^1$-direction, so that the
$g_0$-invariant geodesic line on $\mi (1)$ is the line $\sigma_0=\mi (1) \cap
\{x^2 = 0\}$. The hyperbolic surface $F_0 =\mi (1)/\Gamma_0$ is
homeomorphic to the non-compact annulus $S^1\times \mr$  and its area is not 
finite. The image in $F_0$ of the axis of $g_0$ is a simple 
closed geodesic $c$ of a certain length $s$; give it the positive
weight $r$. So we dispose of a one-component weighted multi-curve
$\mathcal{L}_0$ on $F_0$, as illustrated in figure \ref{fig:tromba}.

\begin{figure}
\begin{center}
\includegraphics[width=12cm]{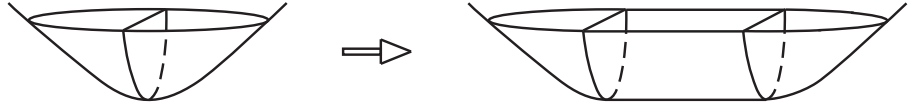}
\caption{Level surfaces $\widetilde{S}_a$ of $U(M_0)$.}
\label{fig:dir}
\end{center}
\end{figure}

\noindent The suspension $M(F_0)= I^+(0)/\Gamma_0$ is a flat spacetime. Let us
now construct $M_0=M(F_0,\mathcal{L}_0)$ which represents the
deformation of $M(F_0)$ associated to the weighted multi-curve
$\mathcal{L}_0$. We shall use the spacetimes $I^+(0)$, $I^+(1,3)$ and
$\Pi^{2+1}$ that we have introduced in section \ref{CTF}. The
universal covering $U(M_0)$ of $M_0$ will be the union of three
domains of $\mm^{2+1}$: $U(M_0) = A \cup B \cup C$, where $ A =
I^+(0)\cap \{x^2 \leq 0\}$, $ B = I^+(1,3) \cap \{0 \leq x^2 \leq
r\}$, $ C = C' + r(0,1,0)$ and $C'= I^+(0)\cap \{x^2 \geq 0\}\ $. In
our notations, $C' + r(0,1,0)$ denotes the set of points in
$\mm^{2+1}$ which can be obtained from $C'$ by means of a translation
of length $r$ along the unit vector $(0,1,0)$.  It is important to
note that the cosmological times of the different pieces $A$, $B$ and
$C$ fit well together; in fact, the CT level surfaces $\widetilde{S}_a$
of $U(M_0)$ are 
$$\widetilde{S}_a =
(a\mi (1)\cap \{x^2 < 0\})\cup (\mi(1,3,a) \cap \{0 \leq x^2 \leq
r\})\cup (a\mi (1)\cap \{x^2 > 0\}+ r(0,1,0))\ .$$
\noindent As shown in figure \ref{fig:dir}, each surface  
$\widetilde{S}_a\subset \mm^{2+1}$ can be obtained by cutting the hyperboloid
$\mi (a)$ along $a\sigma_0$ (which is the intersection of $\mi (a)$ with the
$\{x^2=0\}$-plane) 
and then by inserting a band of $\mi (1,3,a)$
of depth $r\, $.

\noindent The surfaces  $\widetilde{S}_a$
are only $C^1$-embedded into  $U(M_0)$. The initial singularity of
$U(M_0)$ is the segment $J_0=\{\ x^1 = x^3 =0\, \ \ 0\leq x^2\leq r \}$.

\begin{remark}\label{supporting}{\rm We have the following characterization
of $J_0$. The interior points of this segment  make the subset
of $\partial U(M_0)$ (boundary of the convex set $U(M_0)$) of the
points with exactly two {\it null} supporting planes; the end-points make the
subset of $\partial U(M)$ with more than two null supporting planes.
Recall that a supporting plane at $x\in \partial U(M_0)$ is a plane $P$
such that $x\in P$ and $U(M_0)\cap P = \emptyset$. $P$ is null if it contains
some null-lines.}
\end{remark} 

\noindent The covering $U(M_0)\subset \mm^{2+1}$ is flat. 
To get $M_0$, we only need  to specify the action of
$\pi_1=\pi_1(F_0)\cong \mz$ on $U(M_0)$.
\smallskip

\noindent {\bf Action of the fundamental group.} $\pi_1$ acts on $A$ by the
restriction of the  action of $\Gamma_0$ on $\mi (1)$.
The domain $B$ corresponds (via the isometry established in section
\ref{CTF}) to $B'= \{(u,y,\tau)\in \Pi^{2+1}; 0\leq y \leq r\}$,
so that the action of $\pi_1$ on $B$ transported on $B'$ is just given
by the translation $(u,y,\tau)\to (u+s,y,\tau)$. Finally, if $\alpha$
is the translation  $(x^1,x^2,x^3)\to (x^1,x^2+r,x^3)$  on $\mm^{2+1}$,
then the action of $\pi_1$ on $C$ is just the conjugation of $\Gamma_0$ by
$\alpha$.
\medskip

\begin{figure}
\begin{center}
\includegraphics[width=6cm]{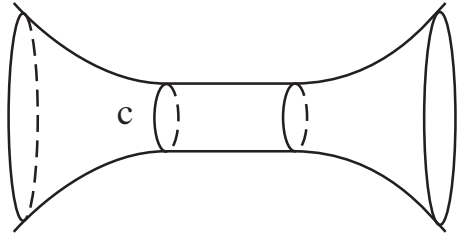}
\caption{Level surface of $M_0$.}
\label{fig:tromba2}
\end{center}
\end{figure}

\noindent The CT of the covering $U(M_0)$ passes to the quotient 
$M_0= U(M_0)/\pi_1$; each level surface
$S_a$ is  only $C^1$-embedded into $M_0$, so that it is endowed
with an induced $C^1$-Riemannian metric.
This allows anyway to define the length of curves traced on the
surface $S_a$  and the derived length-space distance. 
Let $\mathcal{A}= A/\pi_1$, 
$\mathcal{B}=B/\pi_1$ and $ \mathcal{C}= C/\pi_1$. Then, 
$S_a\cap \mathcal{B}$
is a flat annulus of depth $r$ and parallel geodesic boundary
components of length $as\, $; $S_a\cap (\mathcal{A}\cup \mathcal{C})$ can
be isometrically embedded into $aF_0$, and has geodesic boundary
curves of length $as$. As shown in figure \ref{fig:tromba2}, $S_a$ can be
obtained by cutting $F_0$ along $c$ and by inserting a annulus of depth $r$.

\begin{figure}
\begin{center}
\includegraphics[width=12cm]{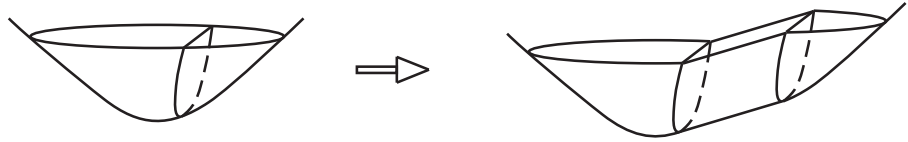}
\caption{Level surfaces of $g(U(M_0))$.}
\label{fig:stor}
\end{center}
\end{figure}

\begin{remark}\label{sigma}{\rm If $g$  is an element of
$ISO^+(2,1)$ acting on $X\in \mm^{2+1}$ as
$g(X) = QX + w$,  the transformed domain $g(U(M_0)) = Q (A\cup B\cup C) +
w$ is, of course, an isometric copy of the universal covering in $\mm^{2+1}$. 
The curve $\sigma = Q(\sigma_0)$ is a geodesic line of $\mi (1)$;
$\sigma$ is the intersection
of $\mi (1)$ with a suitable hyperplane passing at the origin of
$\mm^{2+1}$. Let us denote by $\widehat{\sigma}$ the
suspension of $\sigma \,$; then 

$$Q(B) = \cup_{\lambda \in [0,r]}\ \{\widehat{\sigma} +
\lambda v\}$$ 
\noindent where $v$ is the unitary (in the Minkowski norm)
vector tangent to $\mi (1)$, normal to $\sigma$, and pointing towards
$Q(C')$. We also denote $Q(B) = B(\sigma,v,r)$. The shape of the CT level
surfaces in $g(U(M_0))$ is shown in figure \ref{fig:stor}. 
The initial singularity of  $g(U(M_0))$
is given by  the space-like segment $J= Q(J_0) + w$.}
\end{remark}

\noindent {\bf Simplicial type deformation.} $M_0$ represents a local 
model of the deformation $M=M(F,\mathcal{L})$ we are interested in. In
fact, there exists a neighborhood $\mathcal{W}$ of $\mathcal{B}$ in
$M_0$ which embeds isometrically into $M$, respecting the cosmological
time. Let us denote by $\mathcal{W}'$ the image of $\mathcal{W}$ in $M$.  
Then $M\setminus \mathcal{W}'$ embeds isometrically  into the
Minkowskian suspension $M(F)$, respecting again the cosmological time.

\noindent We describe now the universal covering $U(M)\subset \mm^{2+1}$ and a
cocycle $t:\Gamma \to \mr^3$ which leads to $\Gamma'\subset
ISO^+(2,1)$ such that $M = U(M)/\Gamma'$. The inverse image of
$c\subset F= \mi (1)/\Gamma$ into the covering $\mi (1)$ is an
infinite and locally finite set $\widetilde{\mathcal{L}}$ of disjoint
complete geodesic lines.  Given any geodesic $\sigma_0\in
\widetilde{\mathcal{L}}$, then $\widetilde{\mathcal{L}}= \{\sigma =
\gamma(\sigma_0);\gamma \in
\Gamma\}$. 
Let $\widehat{\mathcal{L}}\subset
I^+(0)$ be the suspension of the geodesic lines of $\widetilde{\mathcal{L}}$. 
The set  $\mi (1)\setminus \widetilde{\mathcal{L}}$ is the union
of an infinite number of connected components.
Denote by $R$ any such a component, and by  $\widehat{R}$ its suspension,
which is a component of $I^+(0)\setminus \widehat{\mathcal{L}}$.
Every $R$ covers a component $F_R$ of $F\setminus \mathcal{L}$;
more precisely, if $\Gamma_R$ is the subgroup of $\Gamma$ which
keeps $R$ invariant, then $F_R = R/\Gamma_R$. 

\noindent Now, fix one base component
$R_0$ and take in it one base point $x_0$. For each $\gamma \in \Gamma$,
let $\gamma (x_0)$ be the point in $\mi (1)$ which is defined by the action
of $\gamma$ on $x_0$. 
The geodesic arc  in $\mi (1)$ connecting $x_0$ with $\gamma(x_0)$ 
crosses a finite number of lines $\{ \sigma_i\}$
belonging to $\widetilde{\mathcal{L}}$. At each crossing consider
the unitary (in the norm of $\mm^{2+1}$) vector $v_i$ tangent to
$\mi (1)$ and normal to $\sigma_i$, pointing far from $x_0$. Then,
the required cocycle $t(\gamma)\in \mr^3$ is given by

$$t(\gamma)= \sum_i rv_i\ .$$

\noindent  Note that if $\gamma_1(x_0)$ and
 $\gamma_2(x_0)$ belong to the same component $R$, 
then $t(\gamma_1)=t(\gamma_2)$,
whence also $t(R)= t(\gamma)$ for any $\gamma$ such that $\gamma(x_0)\in R$, 
is well defined. $U(M)$ is tiled
by tiles of two types: (i) ``$\widehat{R} + w$'',  (ii) 
``$B(\sigma,v,r) + w$'', for some translation vector $w\in
\mr^3$. More precisely, the tiles of the first type make the open
subset of $U(M)$

$$\mathcal{R} = \cup_{R}\ \{\widehat{R} + t(R)\}\ .$$

\noindent Each line $\sigma \in \widetilde{\mathcal{L}}$ is in the
boundary of two regions $R_{\sigma}$, $R'_{\sigma}$ and we assume that
$R_{\sigma}$ is closer to $x_0$ than  $R'_{\sigma}$. Set $v_{\sigma}$ the 
unitary
(in the norm of $\mm^{2+1}$) vector tangent to $\mi (1)$ and normal to
$\sigma$, pointing towards $R'_{\sigma} $. The two regions 
$\widehat{R}_{\sigma} + t(R_{\sigma})$ and  
$\widehat{R}'_{\sigma} + t(R'_{\sigma})$ are connected by the tile of the
second type 
$B(\sigma,v_{\sigma},r) + t(R_{\sigma})$, so that
$$ U(M)\setminus \mathcal{R} = \cup_{\sigma \in
\widetilde{\mathcal{L}}}\ (B(\sigma,v_{\sigma},r) + t(R_{\sigma}))\ .$$
\noindent Note that each tile has its own CT; all the
cosmological times  fit well together and define
the CT of $U(M)$ which passes to the quotient $M$.
\begin{remark}{\rm The construction of $M(F)$ and of $M(F,\mathcal{L})$
can be performed for any hyperbolic surface $F$, not necessarily
compact nor of finite area. Similarly, by starting from {\it any}
locally finite family of weighted geodesic lines in $\mi (1)$, the
simplicial deformation that we have just described produces a globally
hyperbolic spacetime structure on $\mr^2 \times \mr$ with a 
regular cosmological time. }
\end{remark}
\begin{remark}\label{infinitilati}{\rm When $F$ is compact, every 
region $R$ (defined above) is bounded by {\it
infinitely} many lines of $\widetilde{\mathcal{L}}$. In fact, as
$F$ is compact, every $\gamma \in \Gamma$ with  $\gamma \neq 1$  is of
hyperbolic type  \cite{B-P}. Consequently, for any 
$\sigma \in \widetilde{\mathcal{L}}$ and for
every $\gamma \in \Gamma$ with $\gamma (\sigma)\neq \sigma$, 
$\sigma$ and $\gamma (\sigma)$ are ``ultra-parallel''. This means that the
hyperbolic distance satisfies $d(\sigma,\gamma (\sigma))>0$; moreover, 
$\sigma$ and $\gamma (\sigma)$ have a common orthogonal geodesic line. 
Suppose now that a region $R$ is
bounded by  finitely many lines in  $\widetilde{\mathcal{L}}$.
In this case, $R$ contains a band $E$ of infinite diameter, bounded by two
half-lines contained in  $\widetilde{\mathcal{L}}$. As $F$ is
compact, $\mh^2$ is tiled by tiles of the form
$\gamma(\mathcal{D})$, where $\mathcal{D}$ is a fundamental domain
for $\Gamma$ of finite diameter. So, one (at least) tile
$\gamma(\mathcal{D})$  must be contained in $E$. But clearly
$\gamma(\mathcal{D})\cap \widetilde{\mathcal{L}} \neq \emptyset$
and this contradicts the fact that $R$ is a region of $\mh^2
\setminus \widetilde{\mathcal{L}}$.

\noindent The same conclusion holds if $F$ is of finite area. If
$F$ is of infinite area, we can eventually have different
behaviours. For instance, in the example $F_0$ above,
$\widetilde{\mathcal{L}}_0$ just consists of one component which
divides $\mh^2$ into two regions. Other examples will be presented 
in  subsection 4.1.}
\end{remark}

%% file: 4NCOSMO3CT.tex
\section{The cosmological time of (2+1)-spacetimes}
\noindent 
In this section we describe the main properties of the CT for an
arbitrary spacetime $M$. We adopt the notations of the previous
sections; in particular, $M$ is assumed to be an expanding matter-free
spacetime of topological type $S\times \mr$ with compact surface $S$
of genus $g>1$. The validity of the following statements can be quite
easily checked for spacetimes of simplicial type. 
We shall try to
point out the main ideas; we postpone a commentary on the proofs, with
references to the existing literature.

\begin{prop}\label{CTreg} The cosmological time function,
$\tau : M\to (0,\infty[$, is surjective and regular, so that it
defines a global time  on $M$.  It lifts to a regular cosmological
time on the covering, $\widetilde{\tau} : U(M)\to (0,\infty[$. Each
level surface $\widetilde{S}_a$ of $U(M)$ maps onto $S_a$ of $M$ and
is its universal covering. In other words, the action of $\pi_1(S)$ on
$U(M)$ restricts to a free, properly discontinuous isometric action on
$\widetilde{S}_a$ such that $S_a=\widetilde{S}_a/\pi_1(S)$.  Each
$\widetilde{S}_a$ ($S_a$) is a future Cauchy surface.
\end{prop}

\subsection {Initial singularity - external
view}\label{initialsing}

\noindent Let us give a description of the initial
singularity of $M$ as it appears ``from the exterior'' point of view, that
is, from the Minkowski space in which the universal covering $U(M)$ is
placed. In subsection \ref{CTASINTO}
we shall show how the initial singularity can also be
characterized in terms of the observables associated with the
CT asymptotic states.
\smallskip

\noindent Let us first give a definition.
\begin{defi}\label{rt}{\rm An $\mr$-tree (also called a {\it real tree}) is a
metric space $(\mathcal{T},d)$ such that for each couple of points
$p\neq q \in \mathcal{T}$ there exists a {\it unique} arc in
$\mathcal{T}$ with $p$ and $q$ as end-points and this arc is
isometric to the interval $[0,d(p,q)]\subset \mr$. This arc is
called a {\it segment} of $\mathcal{T}$ and is denoted $[p,q]$.}
\end{defi}
\begin{remark}\label{simplicial}
{\rm The so-called {\it simplicial trees} are the simplest examples
of real trees. A  simplicial tree is a real tree covered by a countable family
of {\it elementary} segments, called the ``edges'' of the tree, in
such a way that: a) whenever two edges intersect, then they just have
one common endpoint; b) the edge-lengths take values in a {\it finite}
set of strictly positive numbers. Any endpoint of any edge is called a
``vertex'' of the tree. The distance is the natural length-space
distance. Note that a simplicial tree is not necessarily
locally finite; in other words, vertices of infinite ``valence''
may occur.
In general, a real tree is more complicated than a simplicial tree because
one might find, for instance, a segment containing a Cantor set made by the
endpoints of other segments. }
\end{remark}
\begin{prop}\label{init}
For any $p\in U(M)\subset \mm^{2+1}$ there is a unique time-like geodesic arc
$a(p)$ contained in $U(M)$, which starts at $p$ and is directed 
in the past of $p$, such that the
Lorentzian length of $a(p)$ equals $\widetilde{\tau}(p)$. The
other end-point of $a(p)$, denoted by $i(p)$, belongs to the boundary
$\partial U(M)$ of $U(M)$ in $\mm^{2+1}$. If $p$ and $q$ are
identified in $M$ by the action of $\pi_1(S)$, so are $a(p)$ and
$a(q)$. The union of the initial points 
$\mathcal{T} =\{i(p); p\in U(M)\ \}$ is an 
$\mr$-tree. More precisely, each segment of $\mathcal{T}$ is a
rectifiable space-like curve in $\partial U(M)$ with its own
length. There is a natural isometric action of the fundamental
group $\pi_1(S)$ on $\mathcal{T}$. The quotient space
$i(M)=\mathcal{T}/\pi_1(S)$ can be thought as the initial singularity of
$M$.
\end{prop}
\begin{remark}\label{bestact} {\rm We have already encountered  several 
examples of
spaces  of the form $X=\widetilde{X}/\pi_1(S)$ for some action of
$\pi_1$ on $\widetilde{X}$: for instance,  $F=\mh^2/\Gamma$, $M=U(M)/\Gamma'$,
$S_a = \widetilde{S}_a/\Gamma'$.  Now, the initial singularity of spacetime
also is a quotient
$i(M)=\mathcal{T}/\pi_1(S)$. Instead of the bare topological quotient space,
it  is more interesting  to consider 
$\widetilde{X}$ endowed with the action of $\pi_1$.}
\end{remark}

\begin{figure}
\begin{center}
\includegraphics[width=12cm]{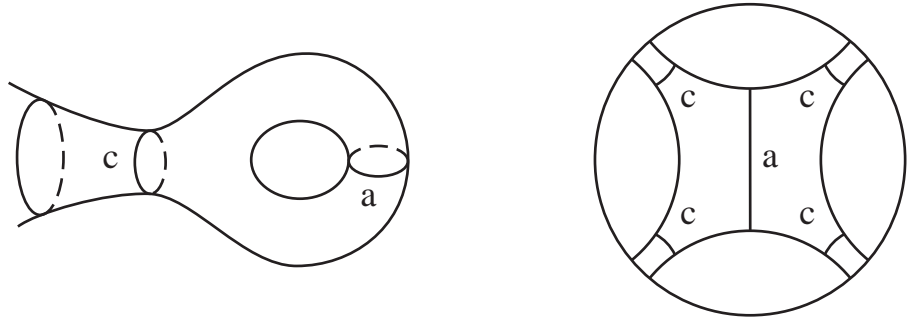}
\caption{Two simple curves on $F=\mh^2/\Gamma$.}
\label{fig:duecurve}
\end{center}
\end{figure}

\begin{figure}
\begin{center}
\includegraphics[width=10cm]{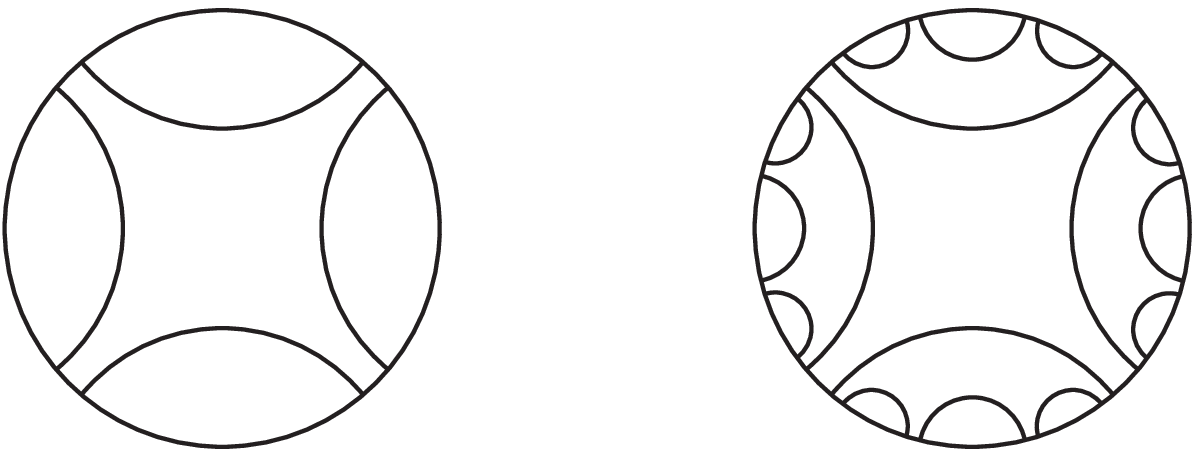}
  \caption{Partial tilings of $\widetilde{F}=\mh^2$.}
\label{fig:partialfund}
\end{center}
\end{figure}

\begin{figure}
\begin{center}
\includegraphics[width=10cm]{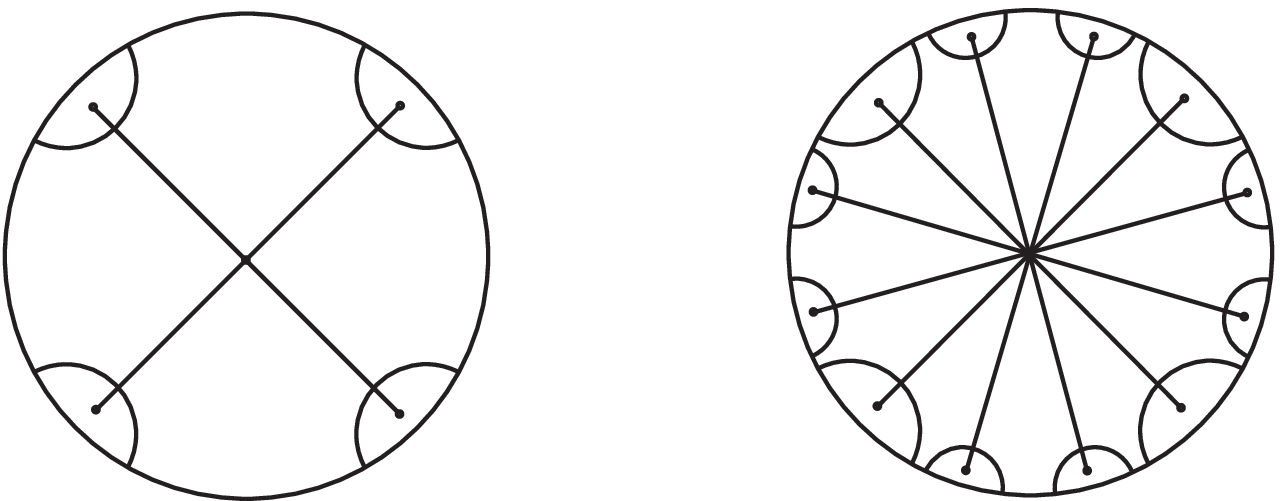}
  \caption{Partial $\widetilde{c}$ and dual trees.}
\label{fig:partialc}
\end{center}
\end{figure}

\begin{figure}
\begin{center}
\includegraphics[width=14cm]{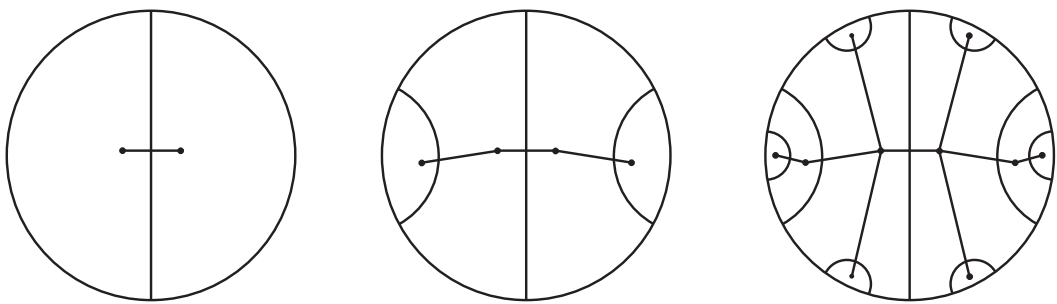}
  \caption{Partial $\widetilde{a}$ and dual trees. }
\label{fig:partiala}
\end{center}
\end{figure}

\begin{remark}\label{simpletype} {\rm When $M$ is of simplicial
type, the corresponding real tree
$\mathcal{T}$ is actually a simplicial real tree. This justifies the
name we have given to these special spacetimes. In this case, the set
of edges of $\mathcal{T}$ consists of the union of the space-like
segments which form the initial singularity of the different tiles of
the form $B(\sigma,v_{\sigma},r) + t(R_{\sigma})$ (see section 3).
The points of $\mathcal{T}$ can also be characterized by the
properties discussed in remark \ref{supporting}.  A homeomorphic (not
isometric) copy of $\mathcal{T}$ can easily be embedded into $\mi
(1)$.  Select one point in each region of $\mi (1)\setminus
\widetilde{\mathcal{L}}$ and consider the set made by the union of all
these points. Connect two points of this set by a geodesic segment of
$\mi (1)$ if and only if they belong to adjacent regions. In this way
we get the required tree. This tree is manifestly ``dual'' of
$\widetilde{\mathcal{L}}$; in fact, the regions of $\mi (1)\setminus
\widetilde{\mathcal{L}}$ correspond to the vertices of $\mathcal{T}$
and the lines of $\widetilde{\mathcal{L}}$ correspond to the edges of
$\mathcal{T}$. We shall return on this duality in section 4.3. Note that, as
demonstrated in remark
\ref{infinitilati}, all the vertices of $\mathcal{T}$ are of
infinite valence.}
\end{remark}
\noindent {\bf Examples of real trees.}
A hyperbolic surface $F=\mh^2/\Gamma$ is represented in figure
\ref{fig:duecurve}. $F$ is of infinite area and is homeomorphic to a
torus with one puncture; two simple closed geodesics $c$ and $a$ on
$F$ are depicted. The geodesic $c$ cuts open $F$ into a compact
surface and an infinite area annulus.  By using the Poincar\'e disk
model, a fundamental domain of $\Gamma$ in $\mh^2$ is also shown in
figure \ref{fig:duecurve}. This domain is delimited by four pair-wise
ultra-parallel geodesic lines. The inverse images of $c$ and $a$ are
represented on this domain. The first two terms of a sequence of partial
tilings of  $\mh^2$, made  by a finite number of copies of the fundamental 
domain, are shown in figure
\ref{fig:partialfund}. The first partial tiling just contains one fundamental
domain. The second is made by the union of $1+4=5$ copies of the fundamental
domain. The next partial tiling of this sequence, which is not shown in the
figure, contains $1+4+12 = 17$ tiles, and so on. For each partial tiling
of $\mh^2$ one can determine a corresponding partial lifting of the
curves $c$ and $a$.  Figure \ref{fig:partialc} shows the first two partial
liftings $\widetilde{c}$ of $c$ and the structure of the associated partial
dual trees. In the limit of the complete infinite tiling of $\mh^2$, the
complete lifting of $c$ contains an infinite number of geodesics and the
associated real tree is infinite. In this case, $\mh^2
\setminus \widetilde{c}$ has exactly one component with infinitely many
boundary lines (the associated vertex of the dual tree has infinite
valence), whereas all the remaining components have one boundary
line. The first three partial liftings $\widetilde{a}$ of $a$ are
shown in figure
\ref{fig:partiala}; the corresponding partial dual trees are also represented. 
Note that these figures are just evocative, as they are not geometrically
exact.

\begin{remark}\label{geometric t}
{\rm The $\mr$-trees and the associated  $\pi_1(S)$-actions which occur in 
proposition \ref{init} are not arbitrary (see \cite{O} pag. 32). 
In fact, one can
prove that the $\pi_1(S)$-action is {\it minimal with small
edge-stabilizers}. This means that there is no non-empty strictly
sub-tree which is invariant for this action, and that, for each segment
in the tree, the subgroup of $\pi_1(S)$ which keeps the segment
invariant is virtually Abelian. We shortly say that a real tree which
admits such a kind of $\pi_1(S)$-action, is {\it geometric}.}
\end{remark}

\subsection {Intrinsic and extrinsic geometry of the
constant CT surfaces}\label{CTsurface}

\noindent In order to describe the geometric properties of the surfaces
of constant cosmological time, it is natural to introduce the notion
of geodesic lamination. 
\begin{defi}\label{gl}{\rm Let $G$ be a surface endowed with a
$C^1$-Riemannian metric. As usual, this induces a length-space
distance on $G$ and the notion of geodesic arc (line) makes sense. A
{\it geodesic lamination} of $G$ is a closed subset $K$ of $G$, also
called the support of the lamination, which is the disjoint union of
complete and simple geodesics, also called the leaves of the
lamination.  ``Complete'' means that we dispose of arc-length
parametrization defined on the whole real line $\mr$;
``simple'' means that the geodesic has no self-normal crossing in $G$.
In other words, each leaf is either a simple closed geodesic or a
simple geodesic which is an isometric copy of $\mr$ embedded in
$G$. When $G$ is compact, such a non compact leaf is not a closed
subset of $G$.}
\end{defi}
\begin{remark}\label{multicur}
{\rm A finite union of disjoint simple
closed geodesics is called a {\it multi-curve} and is the simplest example
of geodesic lamination. We have already introduced multi-curves in
section 3.3. A generic geodesic lamination $K$ can be more complicated than
a multi-curve; in fact, if $\alpha$ is an arc embedded in $G$ which is 
transverse
to the leaves of $K$, typically $\alpha \setminus K$ is a Cantor set.}
\end{remark}

\begin{prop}\label{Sa} For every $a\in (0,\infty[$:

1) $\widetilde{S}_a$ is the graph of a positive convex function
defined on the plane $\{x^3=0\}$ in $\mm^{2+1}$.

2) $\widetilde{S}_a$ is only $C^1$-embedded into $U(M)$,
so that it carries an induced $C^1$-Riemannian metric.
$\widetilde{S}_a$ is geodesically complete and for each $p\neq q
\in \widetilde{S}_a$, there is a unique geodesic arc connecting
$p$ and $q$.

3) The locus  $\widetilde{L}_a$ at which the embedding of
$\widetilde{S}_a$ into $U(M)$ is
no longer $C^2$ is a geodesic lamination of $\widetilde{S}_a$.
$\widetilde{L}_a$ is in fact the
pull-back of a geodesic lamination $L_a$ of $S_a$.
\end{prop}
\begin{remark}{\rm If $M$ is the spacetime of simplicial type which
corresponds to the weighted multi-curve $\mathcal{L}$ on the surface
$F$, then $L_a$ is just made by the boundary components of the flat
annular components embedded into $S_a$, which are associated to the components
of $\mathcal{L}$.}
\end{remark}
\noindent The content of the last remark generalizes as follows.
Recall that $\pi_1(S)$ acts as $\Gamma$ on each $\mi (a)$.
For every  $a\in (0,\infty[ \, $,
let us consider the map
$$p_a : \widetilde{S}_a \to \mi (a)$$

\noindent defined as follows:  $p_a(x)$ is the unique  point
of $\mi (a)$ such that the tangent plane to $\widetilde{S}_a$ at $x$
is parallel to the tangent plane to $\mi (a)$ at $p_a(x)$. This map is
well-defined, surjective and $\pi_1(S)$-equivariant.  By taking the
union of the $p_a$'s we get a $\pi_1(S)$-equivariant map $p: U(M)\to
I^+(0)$, respecting the CT.  This induces a map $p': M\to M(F)$ respecting
the CT.

\begin{prop}\label{p_a}
1) There exists a geodesic lamination $\mathcal{F}$ on $F=\mi
(1)/\Gamma$, which lifts to a geodesic lamination
$\widetilde{\mathcal{F}}$ on $\mi (1)$, such that, for every $a$, one has
$p_a(\widetilde{L}_a)= a\widetilde{\mathcal{F}}$ and any leaf of
$\widetilde{L}_a$ is isometrically mapped 
onto a leaf of $a\widetilde{\mathcal{F}}$. That is, the union of
$p_a(\widetilde{L}_a)$'s covers the suspension $\widehat{\mathcal{F}}$
of $\widetilde{\mathcal{F}}$.

2) $\mathcal{F}$ is the disjoint union of two sublaminations
$$\mathcal{F} = \mathcal{L} \cup \mathcal{F}'$$
\noindent where
$\mathcal{L}$ is the maximal multi-curve sublamination of
$\mathcal{F}$. Note that either $\mathcal{L}$ or $ \mathcal{F}'$
may be empty. Then 

a) $p$ embeds
$U(M)\setminus p^{-1}(\widehat{\mathcal{F}})$ isometrically  
into $I^+(0)$ respecting the CT;

b) $p$ embeds $U(M)\setminus p^{-1}(\widehat{\mathcal{L}})$ continuously
into $I^+(0)$ respecting the CT.

3) The set $p^{-1}(\widehat{\mathcal{L}})$ is the union of components of
the type $B(\sigma,v_{\sigma},r) + w$, so that
$(p')^{-1}(\mathcal{L})\cap S_a$ is the
disjoint union of flat annular components  of $S_a$, like in the case of a
spacetime of simplicial type.
\end{prop}
\noindent We have an immediate corollary concerning the intrinsic and extrinsic
geometry of the constant CT surfaces.
\begin {cor} $\widetilde{W}_a = \widetilde{S}_a \setminus \widetilde{L}_a$
is an open dense set of $\widetilde{S}_a$.
Each component of $\widetilde{W}_a$ is either isometric to an open set of
$\mi (a)$
or is a flat band which embeds into $\mi(1,3,a)$, and projects onto
an annulus of $S_a$. Flat annuli do
occur only if $\mathcal{L}$ is non empty.
\end{cor}

\subsection{ CT duality }\label{CTduality}

\noindent To sum up, two geometric structures are naturally associated 
to the spacetime $M$: the real tree
$\mathcal{T}$ (the initial singularity) and the geodesic
lamination $\mathcal{F}$ on $F=\mi (1)/\Gamma$ which reflects the
lack of smoothness of the embedding of $S_a$ into $M$.
We have already noted that for a spacetime of simplicial type
these two objects are ``dual'' to each other. Here we want to strengthen and 
generalize this point.
\smallskip

\noindent If $\mathcal{L}$ is non empty, we extend the lamination
$L_a$ on $S_a$ to a lamination $L'_a$, by foliating the flat annular
regions of $S_a$ by closed geodesics parallel to the boundary components.
As usually $\widetilde{L}'_a$ denotes the lifted lamination to
$\widetilde{S}_a$. The above map $p_a$ sends  $\widetilde{L}'_a$
onto $a\widetilde{\mathcal{F}}$.

\noindent We have a natural continuous surjective map $i_a:
\widetilde{S}_a \to \mathcal{T}$ which associates to $x$ the initial
point on the arc $a(x)$. So
$\mathcal{T}'=\{i_a^{-1}(x)\}_{x\in \mathcal{T}}$ is a partition of
$\widetilde{S}_a $ by closed subsets. $\pi_1(S)$ acts also on
$\mathcal{T}'$ and, clearly, $i_a$ induces an $\pi_1$-equivariant
identification between $\mathcal{T}'$ and $\mathcal{T}$.

\begin{prop}\label{dual} For every $a$,
each closed set $E$ of the partition $\mathcal{T}'$ of  $\widetilde{S}_a $
is:

1) either the closure of a component of $\widetilde{S}_a \setminus
   \widetilde{L}'_a$;

2) or a leaf of the foliation of some band component of $ \widetilde{L}'_a$
which projects onto a flat annular region of $S_a$ .
\end{prop}

\noindent We describe how the distance $d$ on the real tree $\mathcal{T}$
can be encoded, in dual terms, by equipping the geodesic laminations
$\widetilde{\mathcal{F}}$,
$\mathcal{F}$, with suitable  {\it transverse invariant measures}.
\begin{defi}\label{mgl}{\rm
\noindent A {\it measured geodesic lamination} on $F$ is a couple
$({\cal F}, \mu)$, where $\mathcal{F}$ is a geodesic lamination
and $\mu$ is a {\it transverse invariant measure} which
consists of a {\it Borel measure} $\mu_J$ on each embedded
interval $J \cong [0,1]$ in $F$, transverse to the leaves of $\mathcal{F}$ 
such that

(1) the support of $\mu_J$ coincides with ${\cal F}\cap J$;

(2) if $J,\ J'$ are arcs, homotopic through arcs which are transverse to the
leaves of ${\cal F}$, keeping the endpoints either on the same
leaf or in the same connected components of  $F \setminus {\cal
F}$, then $\mu_J(J) = \mu_{J'}(J')$ . $({\cal F},
\mu)$ lifts to $(\widetilde{\mathcal{F}},\widetilde{\mu})$ which
is $\pi_1$-equivariant.}
\end{defi}
\begin{remark}\label{wmulticur}
{\rm The simplest measured geodesic laminations of $F$
are the weighted multi-curves.}
\end{remark}

\noindent Let $J$ be an arc in $\mi (1)$ transverse to the leaves
of $\widetilde{\mathcal{F}}$. The map $p_a$ lifts $J$ to an arc $J'$
in $\widetilde{S}_a$ transverse to the leaves of
$\widetilde{L}'_a$. On the other hand, by means of the map $i_a$, the
distance $d$ on $\mathcal{T}$ lifts to a measure
$\widetilde{\mu}_{J'}$ on $J'$ which finally gives us the required
($\pi_1$-equivariant) transverse measure on $\widetilde{\mathcal{F}}$.

\noindent One can invert the above construction and associate to
each measured geodesic lamination $({\cal F}, \mu)$ of the
hyperbolic surface $F$ a suitable geometric $\mr$-tree.
\smallskip

\noindent{\bf From geodesic laminations to real trees.} Take the measured 
lamination $({\cal F}, \mu)$ of the surface $F$. $\mathcal{F}$ is in
general the disjoint union of two sublaminations
$$\mathcal{F} = \mathcal{L} \cup \mathcal{F}''$$ \noindent where
$\mathcal{L}$ is the maximal weighted multi-curve sublamination of
$\mathcal{F}$. Note that either $\mathcal{L}$ or $ \mathcal{F}''$ may
be empty. $F\setminus \mathcal{F}$ consists of a finite number of
connected components, the metric completion of any such a component is
isometric to a compact hyperbolic surface with geodesic polygonal
boundary.  If $\mathcal{L}$ is non-empty, let us consider the
spacetime of simplicial type associated to $\mathcal{L}$, and let $F'$
be the $\tau = 1$ level surface of this spacetime. Let us denote by
$\mathcal{F}'$ the lamination on $F'$ which coincides with
$\mathcal{F}$ outside the flat annuli of $F'$ and is defined as $L'_1$
above on these annuli.  If $\mathcal{L}$ is empty, set $F'=F$. The
measured lamination $({\cal F}, \mu)$ ``extends'' to a measured
lamination $({\cal F}', \mu')$ on $F'$. The flat annular components
are foliated by closed geodesics parallel to the boundary
components. These annuli are endowed with a plain transverse measure
of total mass equal to the corresponding annulus
depth. Take the universal covering $\widetilde{F}'$ of $F'$ with the
lifted ($\pi_1$-equivariant) measured geodesic lamination
$(\widetilde{\mathcal{F}'},\widetilde{\mu}')$. Now define a partition
of $\widetilde{F}'$ by closed subsets in the very same way we have
defined above the the partition $\mathcal{T}'$ of $\widetilde{S}_1 $,
with respect to the lamination $\widetilde{L}'_1$. Call again this
partition $\mathcal{T}'$. We can give it a distance $d$ which makes it
an $\mr$-tree. If $E$ and $E'$ are the closure of two components of
the complement of the lamination, take two points $x$ and $x'$ in
these closed sets such that the geodesic segment $[x,x']$ of
$\widetilde{F}'$ is transverse to the leaves of the lamination. By
integration, the transverse measure induces a distance on the subset
of $\mathcal{T}'$ made by the closed sets intersecting $[x,x']$. In
fact, by the ``invariance'' of the measure, this distance doesn't
depend on the segment $[x,x']$. Finally one verifies that in this way
one can actually define a distance between any two points of
$\mathcal{T}'$ and that the resulting $(\mathcal{T}',d)$ is a
geometric real tree.
\begin{remark} {\rm Clearly, weighted multi-curves on the surface $F$ 
dually correspond to
geometric simplicial real trees; the spacetimes of simplicial type do
materialize this duality.}
\end{remark}

\subsection{Reconstruction of
$M=U(M)/\Gamma'$}\label{ricostruzione}

\noindent Starting from  $(F= \mi (1)/\Gamma, \mathcal{T})$ or,
equivalently, from  $(F= \mi (1)/\Gamma,({\cal F}, \mu))$, one can
reconstruct a cocycle $t$, whence $M = U(M)/\Gamma'$. This
generalizes what we have done for a spacetime of simplicial type
in subsection 3.3.
\smallskip

\noindent With the notations introduced at the end of the
previous subsection, consider
$(\widetilde{\mathcal{F}}',\widetilde{\mu}')$ on $\widetilde{F}'$.
To recover a  cocycle $t$ do as follows:
fix one base point $p^*_0$ on $\widetilde{F}'$
out from the support of the lamination. Let $p_0$ be its image on
$F'$. If $\sigma$ is a loop in $F'$ based on $p_0$, which
represents an element $[\sigma]$ of $\pi(F',p_0)$, lift it to the
oriented arc $\sigma^*$ in $\widetilde{F}'$ which starts at
$p^*_0$; up to homotopy we can assume that $\sigma^*$ is
transverse to the leaves of the lamination. Let $f$ be any
continuous $\mr^3$-valued function on $\sigma^*$ which coincides
with the unit normal to the leaves of the lamination, tangent to
$\widetilde{F}'$, and oriented in agreement with $\sigma^*$. Now
we can integrate $f$ along $\sigma^*$ by using the transverse measure
getting a vector $t([\sigma])$. By varying $[\sigma]$, one gets such a 
cocycle $t$.

\subsection {CT asymptotic states}\label{CTASINTO}

\noindent The above discussion tells us that any spacetime $M=U(M)/\Gamma'$
is completely determined by the linear part $\Gamma$ of its holonomy $\Gamma'$
(or equivalently by the surface
$F=\mi (1)/\Gamma$) and by its initial singularity
$i(M)=\mathcal{T}/\pi_1(S)$. The
aim of this subsection  is to recover these geometric objects from the 
``internal point of view'' by
``working inside the spacetime''. More precisely,
 we would like to show  that $F$ and $\mathcal{T}$ can be interpreted as the
future and past {\it asymptotic states} for the geometry of the CT level
surfaces. To this aim we shall consider the observables defined by the
lengths of the curves on the CT level surfaces. It is convenient to introduce
the concept of  
{\it Marked Spectrum} associated with a
metric
space $(\widetilde{X},d)$ which is endowed with an action $\alpha$ of the
surface's fundamental group $\pi_1(S)$, so that 
$X=\widetilde{X}/\pi_1(S)$. Whenever we shall refer to $X$, we
shall actually refer to the triple $(\widetilde{X},d,\alpha)$ 
(see remark \ref{bestact}).

\noindent Let us denote by $\mathcal{C}$ the set of conjugation
classes of $\pi_1(S)\setminus \{1\}$ which coincide with the homotopy
classes of non-contractible continuous maps $f:S^1\to S$.
Each marked spectrum is a
point of the functional space $(\mr_{\geq 0})^{\mathcal{C}}$,
endowed with the natural product topology. The Marked Spectrum
$s_X$ of $X$ (denoted also by $s_{\widetilde{X}}$), is
defined as follows: for any $c=[\gamma]\in \mathcal{C}$, $\gamma
\in \pi_1(S)$, $\gamma \neq 1$,

$$s_X(c)= \inf_{p\in \widetilde{X}}\ d(p,\alpha_{\gamma} (p))\ .$$
\noindent The spectrum is ``marked'' because one takes track
of the map in addition to its image.

\noindent When $X=F$ or $S_a$,   $s_X(c)$ is just the length of a closed
geodesic curve (not necessarily simple; that is, self-crossings
could possibly  occur in $c$) which minimizes the length among the
curves in that homotopy class. For this reason, in such a case,
$s_X$ is called the {\it Marked Length Spectrum} and is denoted by $l_X$.
When $\widetilde{X}=\mathcal{T}$, $s_{\mathcal{T}}$ can
be expressed, in dual terms, as the {\it Marked Measure Spectrum} of the
corresponding measured geodesic lamination $\mathcal{F}$ on $F$;
usually this is denoted $I_{\mathcal{F}}$.
$I_{\mathcal{F}}(c)$ is just the minimal transverse measure
realized by the curves in that homotopy class. When $\mathcal{T}$
is simplicial, that is when $\mathcal{F}$ is a weighted multi-curve
$\mathcal{L}$
of $F$, $I_{\mathcal{L}}(c)$ is easily expressed in terms of the
{\it geometric intersection number} (this also justifies the
notation): assume  that all the weights are equal to $1$ (that is
the length of all edges of $\mathcal{T}$ is equal to $1$); then it
is easy to see that $I_{\mathcal{L}}(c)$ is just the minimum
number of intersection points between $\mathcal{L}$ and any curve
belonging to $c$ and transverse to the components of the lamination.
For arbitrary weights one just takes multiples
of the contribution of each component of $\mathcal{L}$.

\begin{remark}\label{simplecur}{\rm Instead of the whole
$\mathcal{C}$, 
one could prefer to use the
subset $\mathcal{S}\subset \mathcal{C}$ of isotopy classes of {\it
simple} closed curve in $S$,  and take the corresponding (restricted) marked
spectra. The discussion should proceed without any substantial
modification.}
\end{remark}
\medskip

\noindent {\bf On the boundary of the Teichm\"uller space.}
It is convenient, at this stage, to recall the
fundamental facts about the role that the marked spectra
play in the study of the Teichm\"uller space and in Thurston's
theory of its natural boundary. Let us denote by $T_g$ the
Teichm\"uller space of the hyperbolic structures on $S$ 
up to isometry isotopic to the identity. It is 
well-known (see \cite{T3}, \cite{B-P}, \cite {F-L-P}) that the map

$$l: T_g \to (\mr_{\geq 0})^{\mathcal{C}}$$

\noindent defined by $l(F=\mh^2/\Gamma)=l_F$, realizes a 
meaningful embedding of $T_g$ onto a subset of $(\mr_{\geq
0})^{\mathcal{C}}$ homeomorphic to the finite dimensional open ball
$B^{6g-6}$. We shall identify $T_g$ with $l(T_g)$. In fact $T_g$
is in a natural way a real analytic submanifold of $(\mr_{\geq
0})^{\mathcal{C}}$.
\smallskip

\noindent Fix any such a hyperbolic structure $F\in T_g$  on $S$.
Let us denote by $\mathcal{M}\mathcal{G}\mathcal{L}(F)$ the set of
measured geodesic laminations on $F$. Let us denote by
$\mathcal{G}\mathcal{T}(S)$ the set of all $\pi_1(S)$-geometric
$\mr$-trees (remark \ref{geometric t}). At the end of 
subsection \ref{CTduality}, we have outlined a
construction which associates to each $\mathcal{F}\in
\mathcal{M}\mathcal{G}\mathcal{L}(F)$ a dual $\mr$-tree say
$\Delta (\mathcal{F})\in \mathcal{G}\mathcal{T}(S)$. Note that
this construction did not use the fact that $F$ was associated to
a spacetime $M$.
\begin{prop}\label{lam=tree} $\Delta
:\mathcal{M}\mathcal{G}\mathcal{L}(F)\to
\mathcal{G}\mathcal{T}(S)$ is a bijection, that is it can be
naturally inverted. For each $r>0$, $\Delta
(r\mathcal{F})=r\Delta(\mathcal{F}$; here we take either the
$r$-multiple of the measure or the $r$-multiple of the
distance. We can shortly say  that ``$\Delta$ respects the positive
rays''.
\end{prop}
\begin{prop}\label{duality}    Consider the maps,
$ I: \mathcal{M}\mathcal{G}\mathcal{L}(F) \to (\mr_{\geq 0})^{\mathcal{C}}$
and $ s:\mathcal{G}\mathcal{T}(S)  \to (\mr_{\geq 0})^{\mathcal{C}}$,
obtained by taking the corresponding marked
spectra. Then $I = s\circ \Delta$ and is injective. The image in
$(\mr_{\geq 0})^{\mathcal{C}}$ is a positive cone based on the
origin and positive rays go onto positive rays, in a obvious
sense. Moreover $T_g$ and the image Im(I) are disjoint subsets of $(\mr_{\geq
0})^{\mathcal{C}}$
\end{prop}
\begin{remark}\label{observable}{\rm These spectra represent 
the actual ``physical'' observables in our discussion. The last two 
propositions specify the meaning of the duality between laminations and
real trees. As the spectra coincide, they reveal the same physical
content.}
\end{remark}
\noindent Similarly to $T_g$, we identify
$\mathcal{M}\mathcal{G}\mathcal{L}(F)$ and
$\mathcal{G}\mathcal{T}(S)$ with  the image $Im(I)\subset  (\mr_{\geq
0})^{\mathcal{C}}$, endowed with the subspace topology.

\noindent Set $\mathcal{P}^+(\mathcal{M}\mathcal{G}\mathcal{L}(F))=
\mathcal{P}^+(\mathcal{G}
\mathcal{T}(S))=\mathcal{P}^+(Im(I))$ the  projective  quotient
space, obtained by identifying to one point each positive ray in
$Im(I)\setminus \{0\}$. Similarly $T_g\cup \mathcal{P}^+(Im(I))$ has a
natural quotient topology.
\begin{prop}\label{boundary} The pair
$(\bar{T}_g,\partial \bar{T}_g)=(T_g\cup
\mathcal{P}^+(Im(I)),\mathcal{P}^+(Im(I)))$ is homeomorphic to the pair
$(\bar{B}^{6g-6},S^{6g-7})$, where $\bar{B}^{6g-6}$ is the
closed ball and $S^{6g-7}$ is its boundary sphere. The natural
action on $T_g$ of the {\rm mapping class group} $Mod_g$ of $S$
extends to an action on the compactification $\bar{T}_g$. This
is called the Thurston's natural compactification and $\partial
\bar{T}_g$ is the {\rm natural boundary} of the Teichm\"uller
space.
\end{prop}
\noindent We can state precisely how the simplicial trees are
dense, as we claimed in section \ref{GST}. Let us
denote $\mathcal{S} \mathcal{T} (S)$ the subset of
$\mathcal{G} \mathcal{T} (S)$ made by the simplicial real trees.
\begin{prop}\label{dense} $\mathcal{S} \mathcal{T} (S)$ is dense in
$\mathcal{G} \mathcal{T} (S)$ in the induced topology by $(\mr_{\geq
0})^{\mathcal{C}}$.
\end{prop}
\begin{remark}\label{oncompa} {\rm In this remark we collect a few technical
complements
concerning the marked spectra and the geometric meaning of spectra
convergence.

1) The natural compactification of $T_g$ is
formally similar to the
the natural compactification of $\mh^2$ in the hyperboloid model
$\mi (1)$ where $S^1_{\infty}$ is obtained by adding to $\mi (1)$
the endpoints of the rays of the future light cone.

2) Let $F_n$ be a sequence in $T_g$ considered as a sequence of
actions of $\pi_1(S)$ on $\mh^2$. The meaning of the
compactification is the following; up to passing to a subsequence
(still denoted by $F_n$), one of the following
situations occur:  for every $c \in \mathcal{C}$,

(a) $l_{F_n}(c)\to l_{F_0}(c)$, for some $F_0 \in T_g$.

(b) There exist a geometric real tree $\mathcal{T} \in
\mathcal{G} \mathcal{T} (S)$ and a positive sequence $\epsilon_n
\to 0$, such that $\epsilon_nl_{F_n}(c)\to s_{\mathcal{T}}(c)$.
This is also called the Morgan-Shalen convergence of the sequence of
actions. This can be reformulated in a similar, equivalent, dual way
as the convergence (up to positive multiples) of a sequence of
marked length spectra of hyperbolic structures on $S$ to the
marked measure spectrum of a measured geodesic lamination on a
fixed {\it base} $F_0$.

3) The convergence of marked spectra has a deep geometric content.
This can be expressed in terms of the Gromov convergence. Given
two metric spaces $(Y,d)$ and $(Y',d')$ and $\epsilon > 0$, an
$\epsilon$-relation is a set $R \subset Y\times Y'$ (i.e. a
relation between the two spaces) such that:

(i) the two projections of $R$ to $Y$ and $Y'$ are both
surjective;

(ii) if $(y,y'),\ (z,z')\in R$ then $|d(y,z) - d(y',z')| <
\epsilon$.
\smallskip

\noindent Let $G$ be a group, and $\{G\times Y_n\to Y_n\}_{n\geq
1}$ be a sequence of isometric actions of $G$ on the metric spaces
$Y_n$. We say that $(G\times Y_n\to Y_n)\to (G\times Y_0\to Y_0)$
in the sense of Gromov, if for every compact subset $K_0\subset
Y_0$, for every $\epsilon > 0$ and for every finite subset $P$ of
$G$, if $n$ is big enough, there are compact subsets $K_n\subset
Y_n$ and $\epsilon$-relations $R_n$ between $K_n$ and $K_0$ which
are {\it $P$-equivariant}; this means that: if $x\in K_0$, $g\in
P$, $g(x)\in K_0$, $x_n,\ y_n\in K_n$ and $(x,x_n),\ (g(x),y_n)\in
R_n$, then $d_n(g(x_n),y_n) \leq \epsilon$.
\smallskip

\noindent It turns out that in case $(a)$ above we actually have
the convergence in the Gromov sense of the sequence of actions on
$\mh^2$ to an interior point of $T_g$. In case $(b)$, the
Morgan-Shalen convergence is equivalent to the Gromov convergence
for the sequence of actions on $\epsilon_n\mh^2$.

4) Note that $\mathcal{G}\mathcal{T}(S)$ is defined by using only
the topology of $S$ (its fundamental group indeed) while in order
to adopt the dual view point we have to fix (in an arbitrary way)
a base hyperbolic surface $F_0\in T_g$. In fact, the dual view
point can be developed by using the marked measure spectra of the
measured (singular) foliations on $S$ (instead of the measured
geodesic laminations on $F_0$), which only depend on the
differential structure of $S$ (see \cite{F-L-P}).
On the other hand, let us consider $T_g$ as
a space of complex holomorphic structures on $S$ (thanks to the
classical Uniformization Theorem). By fixing any such structure
$F_0$, one can  realize such a spectrum as the measure
spectrum of the horizontal measured foliation of a unique
quadratic  differential $\omega$ on $F_0$. These
``rigidifications'' (via geodesic laminations or quadratic
differentials) of softer objects (the measured foliations) is
reminiscent of the role of Hodge theory with respect to
De Rham Cohomology.}
\end{remark}
\smallskip

\noindent {\bf CT asymptotic states as limit spectra.}
After this somewhat long but necessary digression, let
us come back to the CT asymptotic states. 
\begin{prop}\label{asympt} (a) $\lim_{a\to 0} \ l_{S_a} =  s_{\mathcal{T}}$;
(b) $\lim_{a\to \infty} \  l_{S_a}/a = l_F$.
\end{prop}

\begin{remark}\label{weak effects} {\rm
This means, in particular, that, in a far CT future the spacetime
looks like the Minkowskian suspension $M(F)$. In order to detect the
dual effect of the initial singularity on the embedding of $S_a$ into
$M$, for large value of the cosmological time one needs to increase
the accuracy in the measurement of geometric quantities. Nevertheless
this effect is, in principle, observable for any finite value $a$ of the CT.}
\end{remark}
\begin{prop}\label{analcurve}
For every $a\in (0,\infty[$, $l_{S_a}/a$ belongs to
$T_g\in (\mr_{\geq
0})^{\mathcal{C}}$. Hence, the cosmological time determines a curve $\gamma_M:
(0,\infty[\to T_g$. This is a real analytic curve which connects
$F\in T_g$ with the point on the natural boundary
$[\mathcal{T}]\in
\partial T_g$ (here $[.]$ denotes the projective class).
\end{prop}
\begin{remark}{\rm Consider a
spacetime of simplicial type. To prove proposition \ref{asympt} in this case,
one has to note that the depth of the annular regions is
constant on each $S_a$. When $a\to 0$,  the contribution (to the
length of any curve on $S_a$) of the part contained in the
non-annular components becomes negligible, the length of the
annuli boundaries goes linearly to zero, so that only the
transverse crossing of the annuli becomes dominant. When $a\to
\infty$, the annuli depth goes to zero because of the
rescaling by $1/a$,
and the
length spectrum converge to the spectrum of $F$. The general case
follows by using the density stated in proposition \ref{dense}. 
Concerning proposition
\ref{analcurve}, in the special case of a spacetime of simplicial type, 
the curve in $T_g$ is just
given by the Fenchel-Nielsen flow obtained by ``twisting'' the
hyperbolic surface $F$ along the closed geodesic of the
multi-curve (see \cite{T3} and also \cite{B-P}).}
\end{remark}

\subsection { A commentary on the proofs}

\noindent The identification between
cocycles of a spacetime $M$ with measured geodesic
laminations on $F=\mi (1)/\Gamma$ is due to Mess \cite{Me}. In fact
one can find other examples  of such a construction of ``cocycles''
from measured laminations in the contest of Thurston's theory of
``bending'' or ``earthquakes'' (see for instance \cite{E-M}).
\smallskip

\noindent Measured geodesic laminations emerged in the original
Thurston's approach to the natural compactification of $T_g$
\cite{T} \cite{T2} \cite{T3}. See also \cite{F-L-P} for the
alternative approach by using the measured foliations (see remark
\ref{oncompa} 4). For the claim about the quadratic differentials
in remark \ref{oncompa} 4) see \cite{Ke}. The dual approach via
real trees is due to Morgan-Shalen \cite{M-S} \cite{M-S2}. This
approach does apply to more general, higher dimensional
situations.  The monography \cite{O} contains a rather exhaustive
introduction to this matter and we mostly refer to it (and to its
bibliography) for all the details. In particular one can find in
\cite{O} a complete proof of the duality (see proposition
\ref{lam=tree} and proposition \ref{duality}). The delicate point
is just the inversion of the map $\Delta$ we have described above.
The geometric interpretation of the Morgan-Shalen convergence (see
remark \ref{oncompa} 3) is due to Paulin and Bestvina (c.f. the
bibliography of \cite{O}).
\smallskip

\noindent It is an amazing fact that the spacetimes
``materialize'' this subtle duality in the way we have seen. Note
also that, in the spacetime setting, the
choice of the base hyperbolic surface $F$  (see remark
\ref{oncompa} 4) is fixed  by the linear part of the
holonomy of $M$, that is by its future asymptotic state.

\noindent Concerning proposition \ref{analcurve}, the
Fenchel-Nielsen flow generalizes to the earthquake flow (one uses
again the density \ref{dense}) with initial data $(F,\mathcal{F})$
which has real analytic orbits \cite{Ke2}.

%% file: 4NCOSMOCOM.tex
\section{Complements}
\noindent In this section we add a few comments  
about the flat spacetimes
with compact space of genus $g=1$, and about the spacetimes with
negative cosmological constant. Finally we discuss a conjecture relating
the CT and the CMC time.

\subsection{Toric space ($g =1$)}

\noindent The case in which the surface $S$ is a torus
is a particular example of the so called {\it Teichm\"uller
spacetimes} which we have analysed in \cite{BG}. So we simply
remind the main points. Each non static spacetime determines a curve 
$\gamma:(0,\infty[\to T_1^*$, $ \gamma(a)=(w(a),\omega(a))$, where $T_1^*$ is 
the cotangent bundle of the Teichm\"uller space $T_1$ of conformal structures
on the torus up to isomorphism isotopic to the identity. Let us recall that
$T_1$ is isometric with the Poincar\'e disk. 
The cotangent
vectors $\omega(a)$ at the point $w(a)\in T_1$ is a quadratic differential
on a Riemann surface representing $w(a)$.
It is not hard to verify that $\gamma$  is just the complete
orbit of the Teichm\"uller flow with initial data
$(w(1),\omega(1))$ (see \cite{Ab},\cite{BG}). In particular the
projection of $\gamma$ onto $T_1$ is a complete geodesic
connecting two boundary points. These points can also be 
interpreted in terms of marked spectra. Let us denote by
$\mathcal{H}$ and by $\mathcal{V}$ the horizontal and vertical
measured foliations of $w(1)$. Then: $\lim_{a \to \infty}\ l_{S_a} /a =
I_{\mathcal{H}}$ and $\lim_{a \to 0}\ l_{S_a} = I_{\mathcal{V}}$.

\subsection{Spacetime with negative cosmological constant}

\noindent The above discussion on  CT for flat spacetimes (i.e.
with cosmological constant $\Lambda =0$) can be adapted to the case of
negative $\Lambda$ which we normalize to be $\Lambda = -1$. We denote
by $\mx^{2+1}$ the Universal anti de Sitter Spacetime of dimension
$2+1$. Each spacetime is now locally isometric to $\mx^{2+1}$. The
role played by $I^+(0)$ in the flat case, is played now by the
diamond-shaped domain $D(2)$ (see \cite{H-E} pag. 132) isometric to
$B^2\times (-\pi /2,\pi /2)$ with metric, in coordinates $(y^1,y^2,\
t)$, $ ds^2 = (\cos^2t) h_2 -dt^2 \ $, where $h_2$ is the usual
Poincar\'e hyperbolic metric on the open disk $B^2$.
\medskip

\noindent {\bf Anti-de Sitter Suspensions.} If $F = \mh^2/\Gamma$ is a
hyperbolic surface of genus $g>1$, then $\Gamma$ isometrically
acts also on $D(2)$ and $ D(F) = D(2)/\Gamma$ is the anti-de
Sitter suspension of $F$. Up to a translation, the function $t$
gives the CT and it has many qualitative properties similar to the
CT of the Minkowskian suspensions, but we have now both an initial
and a final singularity, both reduced to one point. In a sense,
$D(F)$ can be  obtained by the Minkowskian suspension $M(F)$ by a
procedure of {\it warping and doubling}; $D(F)$ and $M(F)$ have the
same initial singularity; the future asymptotic state of $M(F)$
``becomes'' the level surface of the CT on $D(F)$ where the
expansion ends and the collapsing begins. Also the anti-de Sitter
analogous of $I^+(1,3)$ is easy to figure out.
\medskip

\noindent {\bf Deforming anti-de Sitter suspensions.}
We want to generalize the above ``warping and doubling''
construction. Let $M = U(M)/\Gamma'$ as in the former
flat-spacetime discussion, $\Gamma' = \Gamma + t(\Gamma)$. $F=\mi
(1)/\Gamma$ as usually. For $t\in (-\pi /2,0)$, $\tau \in
(0,\infty)$, set $t = -(\pi /2) e^{-\tau}$. Denote $h(a)$ the
spatial metric on $S_a$. On the manifold $F \times (-\pi /2,0)$
consider the metric $ ds^2 = \cos^2(t) h(\tau)/\tau^2 - dt^2\ $,
getting a spacetime $\mathcal{D}'(M)$. Similarly, take $M^-$ and
$-\mathcal{D}'(M^-)$, where  $M^- = U(M^-)/-\Gamma'$,
$-\Gamma'=\Gamma -t(\Gamma)$, $-\mathcal{D}'(M^-)$ is obtained
from $\mathcal{D}'(M^-)$ by reversing the time and the
orientation. Finally set $\mathcal{D}(M) = \mathcal{D}'(M)\cup
-\mathcal{D}'(M^-)$, by gluing the two pieces at $t=0$.
$\mathcal{D}(M)$ is locally anti de Sitter; up to a translation,
$t$ gives the CT. The asymptotic state for $t\to -\pi /2$ (i.e.
the initial singularity) is equal to the initial singularity of
$M$. The final singularity ($t\to \pi /2$) coincides with the
initial singularity of $M^-$. The future asymptotic states of $M$
and $M^-$ ``glue'' at the level surface $\{t=0\}$ of the CT  where the
expansion ends and the collapse begins. The orbit of
$\mathcal{D}(M)$ in $T_g$ is given by the union of two earthquake
rays associated to $M$ (pointing to the future) and to $M^-$
(towards the past); note that the qualitative behaviour is similar
to what we have remarked for $g=1$. $\mathcal {D}(M)$ is the
quotient of a domain $D(2)_M\subset \mx^{2+1}$, which is a
``deformation'' of the diamond-shaped domain $D(2)$. Also in this
case the spacetimes with simplicial asymptotic singularities are
significant and particularly simple to be understood.

\subsection{ CT versus CMC}
\noindent Assume again that the space $S$ is of genus $g>1$, and
that the spacetimes are flat. Given any global time
on a spacetime $M=U(M)/\Gamma'$, the asymptotic behaviour of the geometry
of the corresponding level surfaces reflects in general a property of the
specific time and not of the spacetime. On the other hand, we have seen that
the asymptotic states of the cosmological time are intrinsic features of the
spacetime. In this sense, we can say that a global time is ``good'' 
when it has the
same asymptotic states of the CT. The CMC time, $\rho$ say, is a widely studied
global time. A natural question is whether $\rho$ is a good global time. 
We conjecture that this is the case.
Let us denote by $W_a$ the  $\{\rho =a\}$ level surfaces of the CMC time.
\begin{conge} (a) $\lim_{a\to \infty}\ l_{W_a} = s_{\mathcal{T}}$;
(b) $\lim_{a\to 0}\ l_{W_a}/a  =  l_F$.
\end{conge}

\noindent There are some strong evidences supporting the
conjecture; in particular by \cite{A-M-T} we know that:

(1) $\rho$ is a global time function with image the interval
$(0,+\infty)$.

(2) If $ \gamma: (0,\infty) \to T_g$ is any $\rho$-orbit in $T_g$
(here $T_g$ is intended as a space of conformal structures) then:

(i) The $\lim_{\rho\to 0}\ \gamma$ exists in $T_g$.

(ii) $\gamma$ is proper, that  is it goes out from any compact set
of $T_g$, roughly it ``goes to $\infty$''.

\noindent An idea to prove the conjecture, should be to
confine each $W_a$ between two barriers made by  CT-level surfaces
$S_{a'}$, $S_{a''}$, in such a way that $a'$ and $a''$ depend nicely
on $a$ and, when $a\to \infty$ or $a\to 0$, $S_{a'}$ and $S_{a''}$
become more and more ``geometrically'' close to each other. In a
recent conversation, L. Andersson confirmed that this should
actually work at least for a spacetime with simplicial initial
singularity. A similar conjecture can be formulated in the anti-de
Sitter context.